\documentclass[aps,prb,preprint,groupedaddress,citeautoscript,nopacs]{revtex4}

\usepackage{graphicx}
\usepackage{color}
\usepackage{amsfonts}
\setcitestyle{super}

\begin{document}
\title{ 
\begin{flushright}\vspace{-1in}
       \mbox{\normalsize  EFI-14-30}
       \end{flushright}
Non-Fermi Liquid in Dirac Semi-metals
}
%

\author{Eun-Gook Moon$^{1, 2}$ and Yong Baek Kim$^{3}$}

\affiliation{$^1$ Kadanoff Center for Theoretical Physics and Enrico Fermi Institute, University of Chicago, Chicago, IL 60637, USA}

\affiliation{$^2$ Department of Physics, University of California, Santa Barbara, CA 93106, USA}

\affiliation{$^3$ Department of Physics,University of Toronto, Toronto, Ontario M5S 1A7, Canada}

\date{\today}

\begin{abstract}
{\bf 
Quantum criticality,  a manifestation of emergent scale invariance in electron wavefunctions arises from intricate many-body quantum entanglement. 
One of the natural venues for the criticality is clean undoped Dirac semimetals, known as a marginally-renormalized critical phase.
The ground state is only slightly modified from the Slater-type product wavefunction because the scatterings from weak disorder and Coulomb interactions are suppressed.
Here, using the renormalization group (RG) analysis, we show that a novel class of quantum criticality 
appears in Dirac semimetals when the disorder strength becomes sufficiently strong in the presence of 
Coulomb interactions so that a quantum phase transition from the marginally-renormalized critical phase to 
a disorder-dominated phase arises. 
The ground state at the critical point is a quantum critical {\em non-Fermi liquid} (NFL),
characterized by the properties of strongly entangled low-energy states such as the absence of quasi-particles. 
Near the critical point, unusually wide temperature regions with NFL behaviors emerge.
}
\end{abstract}

\maketitle

Discovery and investigation of quantum criticality have been important paradigms for fundamental
understanding of many-body correlations in condensed matter systems.
Coulomb interactions and disorder are two impetuses of quantum criticality \cite{subir, localization}.
The former usually drives spontaneous-symmetry-breaking as in magnetism, superfluidity, and superconductivity while 
the latter gives rise to metal-insulator transitions between localized and delocalized states.
In spite of extensive research on quantum criticality,  the results of interplay between the two impetuses 
still remain illusive in most cases.

In Dirac semi-metals \cite{dirac1, dirac2, dirac3, dirac4, dirac5, dirac6, dirac7, dirac8, dirac9},
the low-energy spectrum of the non-interacting electrons is scale-invariant and has a finite number of Fermi points in the Brillouin zone 
with a linear dispersion relation. The density of states has the power-law form, $\mathcal{D}(E) \sim |E-E_F|^2$ ($E_F=0$ from now on).
The small density of states represses the scatterings from weak disorder 
in sharp contrast to conventional Fermi liquids \cite{fradkin1, fradkin2, shindou, gurarie, goswami, sarma, biswas, herbut} 
and induces a marginally-renormalized quantum critical phase from long-range Coulomb interactions \cite{hosur, isobe}. 
Thus, the ground state is only slightly modified from the Slater-type product wavefunction.

In this work, we unveil a novel class of quantum criticality driven by the interplay between the Coulomb interactions and disorder in Dirac semi-metals.
This quantum criticality is associated with a quantum phase transition driven by varying strength of disorder, 
between a disorder-dominant ground state and the marginally-renormalized quantum critical phase.
Remarkably, the ground state at the critical point is a quantum critical {\em non-Fermi liquid}. 

Striking properties arise in the NFL and associated finite-temperature regions.
Physical quantities are characterized by power-law correlations with different critical exponents from ones of the marginally-renormalized Dirac semimetals.
As a result, their behaviors are qualitatively different from the corresponding quantities of the marginally-renormalized Dirac semimetals. 
For example, the density of states ($\mathcal{D}(E) \sim|E|^{3/(1+\frac{\epsilon}{2}) -1}$) near the critical point ($\epsilon=1$ for the white noise disorder) is enhanced from that of the weakly interacting ground state ($\mathcal{D}(E) \sim|E|^{2}$ { up to} logarithmic corrections). 
Moreover, quasi-particles are not well-defined with the anomalous dimension, $\eta_f$ (introduced below), for the electron spectral function, and the finite-temperature quantum-critical region is unusually wide as illustrated in Fig. 1.

Systematic understanding in  the interplay between the Coulomb interaction and disorder is achieved by generalizing the white-noise quenched disorder to spatially-correlated one.
The generalization enables us to control RG calculation with a small parameter and 
understand the underlying physical mechanisms behind the NFL behaviors;
the competition and balance between disorder-driven enhanced density-of-states and  the screening of the Coulomb interactions by particle-hole excitations. 
Such competition is a driving force { for} NFL, which also appears in the interacting electron systems with quadratic band-touching
in three-dimensions \cite{moon,herbut2}.

\section*{Model and basic scaling properties}

We consider the kinetic Hamiltonian of the Dirac semi-metals,
\begin{eqnarray}
\mathcal{H}_0 (k) = \sum_{a=1}^3 \epsilon_a(k) \Psi^{\dagger} \Gamma^a \Psi, \quad \epsilon_{1,2,3}(k) = v_F k_{x,y,z}.
\end{eqnarray}
$\{ \Gamma^a \}$ denote mutually anti-commuting matrices whose minimal dimension is 4 for one nodal point. 
Energy spectrum is $E_{\pm}(k) = \pm v_F \sqrt{k_x^2+k_y^2+k_z^2}$. For simplicity, we assume an isotropic velocity, which is valid at low energy with the long-range Coulomb interactions.

The long-range Coulomb interactions in pure (disorder-free) Dirac semimetals are well understood in literature \cite{hosur, isobe}.
The Coulomb interaction, $H_{C} = \int_x \int_y n(x) n(y)V(|x-y|)$ with $n(x)= \Psi^{\dagger}(x)\Psi(x)$ and $V(r)=e^2 /(2r)$, becomes marginally irrelevant, which is captured by the beta function of $g_1\equiv\frac{2 e^2}{ \pi v_F}$,
\begin{eqnarray}
\frac{d}{dl} g_1 = - (N_f \,c_{\varphi}+\frac{1}{3})g_1^2.
\end{eqnarray}
The number of Dirac points, $N_f$, and $c_{\varphi} = \frac{ {\rm Tr} (\Gamma^0) }{12}$ are introduced. 
Notice that $l \rightarrow \infty$ is the infrared limit. 
The minus sign of the right hand side indicates that the Coulomb interaction is screened by particle-hole excitations. 
This flow equation suggests that the non-interacting ground state is marginally renormalized by the long-range Coulomb interactions,
only with logarithmic corrections.

Quenched disorder effects are captured by the impurity scattering Hamiltonian , ${H}_{imp} = \int_x r_{\kappa} \, \eta(x) \Psi^{\dagger}M^{\kappa}\Psi(x)$.
The matrices $M^{\kappa}$ represent different types of disorder. 
We focus on the most important one, chemical potential disorder $M^0= \Gamma^0 = I$, and less important types of disorders are discussed in the supplementary information.
The coupling constant $r_{0}$ controls the strength of disorder. 
Disorder properties are further characterized by the following correlation functions,
\begin{eqnarray}
\langle \eta(-q,\tau) \eta(q,0) \rangle = \frac{1}{q^m} \rightarrow \,\,\langle \eta(x,\tau) \eta(0,0) \rangle \sim \frac{1}{|x|^{3-m}}, \nonumber
\end{eqnarray}
with $\langle \eta(x,\tau) \rangle =0$. 
The bracket indicates disorder ensemble average and the arrow is for the Fourier transformation.  
$m=0$ corresponds to the white noise disorder, and a non-zero $m$ describes spatially-correlated quenched disorder.

Let us first consider a mean field theory for the relaxation time in the non-interacting electron system 
with the white-noise disorder ($m=0$ and $e^2=0$). The self-consistent Born approximation gives
\begin{eqnarray}
\frac{1}{\tau} \equiv {\rm Im} \Sigma_{Born} (0, \omega=0) = r_0^2 \int^{\Lambda} \frac{d^3 k}{(2\pi)^3} \frac{\frac{1}{\tau}}{E(k)^2+\frac{1}{\tau^2}}. \nonumber
\end{eqnarray}
The momentum integration is ultraviolet divergent and regularized by the cutoff $\Lambda$.
The critical value of the disorder strength is then determined by 
$1=r_{0c}^2 \int^{\Lambda} \frac{d^3 k}{(2\pi)^3} \frac{1}{E(k)^2}$, 
and the mean field solution is
\begin{eqnarray}
\frac{1}{\tau} = 4 \pi\Big(\frac{1}{r_{0c}^2}-\frac{1}{r_{0}^2} \Big) \Theta(\frac{r_0}{r_{0c}}-1). \nonumber
\end{eqnarray}
The mean field theory describes a phase transition between the non-interacting ground state and a disorder-dominant ground state \cite{fradkin1, fradkin2}.

To go beyond the mean field treatment, we consider the action with disorder-fluctuation and Coulomb-potential fields. 
\begin{eqnarray}
\mathcal{S} &=& \int_{x,\tau}\psi^{\dagger}_{\gamma}\Big[ \partial_{\tau}+ \mathcal{H}_0(-i \nabla) - r_{\kappa} \, \eta M^{\kappa} -i e \varphi_{\gamma} \Big]\psi_{\gamma} \nonumber \\
&&+ \frac{1}{2} \eta \Pi_\eta \eta + \frac{1}{8\pi} (\partial \varphi_{\gamma})^2,
\end{eqnarray}
where $\varphi \,(\eta(x,\tau))$ represents the Coulomb-potential (disorder-fluctuation) fields. 
The index, $\gamma=1,\cdots,R$ is introduced for the replica trick and implicit below. 
Integrating out the Coulomb-potential and disorder-fluctuation fields leads to the $1/r$ potential and disorder-induced couplings.
For example, the white noise disorder has $\Pi_\eta(q,\omega_n) = \delta(\omega_n)$ (suppressing finite frequency compoenents of the $\eta$ field), and integrating out the $\eta$ field leads to
$\frac{-r^2}{2}\int_x \int d\tau d\tau{'}  \psi^{\dagger}(x,\tau)M \psi(x,\tau)  \psi^{\dagger}(x,\tau{'})M \psi(x,\tau{'})  $.

The scaling analysis with a generic dynamic critical exponent,
$\tau \rightarrow \tau e^{-z l}$ and $x_i \rightarrow x_i e^{- l}$, gives the scaling dimensions of operators and coupling constants,
$ [\psi^{\dagger}\psi]= 3$, $[\eta] = \frac{3 -m}{2}$, $[r_b] = z- \frac{3-m}{2}$, and $[e^2]=z-1$. 
Note that the contact interactions $\int_{x,\tau} (\psi^{\dagger} M \psi)^2$ are irrelevant, so we only focus on the long-range one. 

For the non-interacting Dirac semi-metals, it is natural to assign the dynamic critical exponent 
$z=1$. Then,  $[r]=\frac{m-1}{2}$ and $[e^2]=0$, so the disorder becomes marginal when $m_c = 1$. 
As expected in the mean field discussion, the white noise disorder ($m=0$) is irrelevant for the non-interacting ground state. 
To control our calculation, we introduce $\epsilon =1-m$ and use it for the expansion parameter. 
The white noise limit is taken $\epsilon \rightarrow 1$ at the end of calculations.

Perturbative computation of the electron self-energy for the disorder scattering (Fig. \ref{diagram} (a) the first diagram) leads to
\begin{eqnarray}
\Sigma_{f}(k, i \omega) &\equiv& \Sigma^{\alpha}_f(k, i \omega) \Gamma^{\alpha}= r_{0}^2\int_{q} G_{f}(k+q, i \omega_n) \frac{1}{\Pi_{\eta}} ,\nonumber 
\end{eqnarray}
where $\alpha=0,1,2,3$. 
The electron Green's function and the disorder-field propagator are defined as $G_f^{-1}(k, i \omega_n) = -i \omega_n + \mathcal{H}_0(k) $, and $\Pi^{-1}_{\eta}$.
The zeroth component is
\begin{eqnarray}
\Sigma_f^0(k, i \omega_n) =r_{0}^2 \int_q \frac{i \omega_n}{\omega_n^2 + E(k+q)^2} \frac{1}{q^m}, \nonumber
\end{eqnarray}
with $E(k)^2 = \sum_a \epsilon_a^2(k)$.
The vertex correction (Fig. \ref{diagram} (c) the first diagram) is 
$V_{\kappa} = \int_{q} \frac{1}{E_q^2}\frac{r_{0}^2}{q^m}$.
The positiveness of the vertex correction indicates that the chemical potential disorder effect is enhanced 
via disorder scatterings (See supplementary information for detailed calculation).

The relaxation rate is obtained from the imaginary part of $\Sigma^0(0, \omega + i \eta)$ after analytic continuation, 
\begin{eqnarray}
\frac{1}{\tau(\omega)} = {\rm Im} \Sigma^0(0, \omega ) \sim r_{0}^2 \, \omega^{1+\epsilon}. \nonumber
\end{eqnarray}
The suppressed scattering rate at low energy is consistent with the necessity of the critical strength of disorder to drive
a phase transition, as shown in the mean field calculation, which is in sharp contrast to a constant scattering rate in Fermi liquids.

\section*{Renormalization group analysis}

Let us first consider the renormalizaton group (RG) flow for the case without the Coulomb interactions, $e^2=0$.
To characterize the strength of chemical potential disorder, we introduce a dimensionless parameter, $
g_{2} = \frac{r_{0}^2 \Lambda^{\epsilon} }{2 \pi^2 v_F^2 }$, with a large momentum cutoff, $\Lambda$.
When $\epsilon=1-m = 0$, the relevant Feynman diagrams are logarithmically divergent, so it is straightforward to identify the divergent pieces. 
For example, the divergent piece of the zeroth component of the self-energy is given by
\begin{eqnarray}
\Sigma_f^0(k, i \omega_n)|_{div} = (i \omega_n) g_{2} \log \Lambda. \nonumber
\end{eqnarray} 
Along the same line, the vertex correction and other self-energy components also have logarithmic divergences, which can be straightforwardly analyzed (see supplementary information for details).
Putting these contributions together leads to the following beta function.
\begin{eqnarray}
&&\frac{d}{d l} g_{2} = - \epsilon \,g_{2} + \frac{8}{3} g_{2}^2.
\end{eqnarray}
There exist two fixed points $g_{2} =(0, \infty)$, which correspond to a non-interacting ground state and a disorder-dominant state. 
For the disorder-dominant state, one can expect either a diffusive metallic state or a localized state, but 
our calculation cannot {\it a priori} pin down the nature of the ground state with strong disorder. 
Previous works suggest that the disorder-dominant phase is a diffusive metallic state \cite{fradkin1, fradkin2, shindou, gurarie, goswami, biswas, herbut,sarma}.
The unstable fixed point $g_{2} =\frac{3\epsilon}{8}$ describes a quantum phase transition between the two stable fixed points. 
It is clear that the $\epsilon$ expansion controls the critical point. 
Physical implication of these fixed points is further discussed below after accounting for the long-range Coulomb interactions.

Now we turn on the electric charge to understand the interplay between disorder and the Coulomb interactions.  
Taking into account the contributions from the remaining five Feynman diagrams in Fig. \ref{diagram} (see supplementary information), 
the beta functions of the two coupling constants are obtained as 
\begin{eqnarray}
&&\frac{d}{d l} g_1 = - (N_f c_{\varphi}+\frac{1}{3}) (g_1)^2+\frac{4}{3} g_1 g_{2} \nonumber \\
&&\frac{d}{d l} g_{2} = - \epsilon \,g_{2} + \frac{8}{3} (g_{2})^2 -\frac{2}{3} g_1 g_{2}, \label{beta}
\end{eqnarray}
and the renormalized dynamic critical exponent is $z=1- \frac{g_1}{3}+\frac{4 g_2}{3}$ setting the velocitiy to be a constant. 
The RG flow diagram is illustrated in Fig. \ref{flowdiagram}. There are three fixed points, $(g_1, g_{2})= \big[(0,0), (0, \frac{3 \epsilon}{8})$, $ (\frac{\epsilon}{2 N_f c_{\varphi}}, (3+\frac{1}{N_f c_{\varphi}}) \frac{\epsilon}{8})\big]$ for non-zero positive $\epsilon$ in addition to the 
disorder-dominant fixed point ($D$).

The marginally-renormalized Dirac semi-metal fixed point $O =(0,0)$, where the Coulomb interaction is marginally irrelevant, 
is stable in the presence of both disorder and Coulomb interaction.
The beta function for $g_2$ suggests that disorder is irrelevant for positive $\epsilon$, so it is legitimate to use
perturbative analysis.
The disorder scattering rate obtained earlier, $\frac{1}{\tau(\omega)} \sim \omega^{1+\epsilon}$, can be used up to 
small corrections from irrelevant perturbations.
The ground state is then a quantum critical phase, where the Dirac semi-metal is weakly modified by the marginally 
irrelevant Coulomb interactions with logarithmic corrections in various physical quantities.
{ We call this phase as the} marginally-renormalized quantum critical phase, as illustrated in Fig. \ref{phasediagram}. 

At $U=(0,\frac{3 \epsilon}{8})$, 
both disorder and the Coulomb interactions are relevant. 
The relevance of the long-range Coulomb interactions can be understood from the renormalized dynamic critical exponent, $z= 1+\frac{\epsilon}{2}$ and the scaling dimension of the Coulomb interactions, $[e^2]_{U}=(z-1)= \epsilon/2$.
Note that the larger dynamic critical exponent induces more density of states ($\mathcal{D}(E) \sim |E|^{3/z -1} $), which leads to enhanced Coulomb interactions.

As illustrated in Fig. 3, there is a new fixed point with one relevant operator, $ QC=(\frac{\epsilon}{2 N_f c_{\varphi}}, (3+\frac{1}{N_f c_{\varphi}}) \frac{\epsilon}{8})$. 
The RG-flow out of $U$ goes into $QC$, which describes a quantum phase transition between the marginally-renormalized 
quantum critical phase ($O$) and the disorder-dominant ground state ($D$). 
Note that the presence of this fixed point is controlled at $O(\epsilon)$.
The appearance of the fixed point is a consequence of the interplay between the enhanced density of states 
and the screening of the Coulomb interaction.
Near $U$, the RG-flow is governed by the enhanced density of states, but later the screening of the Coulomb interaction by
particle-hole excitations balances out this effect and the RG flow finally leads to the fixed point $QC$.

The fixed point $QC$ determines the properties of the quantum critical point for the transition between the marginally-renormalized
critical phase ($O$) and the disorder-dominant phase ($D$).
The ground state at the quantum critical point is a NFL; all physical quantities receive anomalous dimensions except for 
the number operator protected by the Ward Identity. 
We emphasize that the ground state is a {correlated}  many-body state with non-zero Coulomb interactions, $g_1|_{QC} \neq 0$.
The electron anomalous dimension is $\eta_f = (3+\frac{1}{N_f c_{\varphi}}) \frac{\epsilon}{8}$ with the renormalized dynamic critical exponent $z = 1+\frac{ \epsilon}{2}$.
Thus, the electron spectral function has a branch-cut structure,
\begin{eqnarray}
A(k, \omega) = {\rm Im} G_f (k, \omega) \sim \frac{\Theta (\omega-c_m k^z )}{|c_m k^z - \omega|^{1-\eta_f}},
\end{eqnarray}
where $c_m$ is a non-universal dimensionful constant and $\Theta(x)$ is the step function. 

In addition to the electron spectral function, temperature dependence of the specific heat is $C_{v}(T) \sim T^{3/z}$ that is related to the renormalized density of states, $\mathcal{D}(E) \sim E^{3/z-1}$. 
Optical conductivity has the scaling behavior, $\sigma(\omega, T) \sim \omega^{1/z} \mathcal{F}(\frac{\omega}{T})$. 
Note that the conductivity shows more ``metallic'' behaviors than that of the non-interacting ground state ($\sigma(\omega) \sim \omega$). 
Scaling functions of physical operators with full symmetry classification of electron bilinear operators will be presented in another place.
The crossover boundary of Fig. \ref{phasediagram} is determined by comparing scaling dimensions of temperature and the relevant operator $g_2$ at $QC$, $T^* \sim |g_2 -g_{QC}|^{z\nu}$, where $\nu^{-1}$ is the positive eigenvalue of the linearized beta function, $ \nu^{-1} = (3+(N_f c_{\varphi})^{-1} + \sqrt{81+30/(N_f c_{\varphi}) + 1/(N_f c_{\varphi})^2}) \frac{\epsilon}{12}$. One can easily show that $z \nu > 1$ 
that allows unusually wide critical fan in Fig. 1.	
The electric potential also receives the anomalous dimension, $\eta_{\phi} = \frac{\epsilon}{2}$, which is another signature of the quantum critical NFL \cite{moon}.
The Coulomb potential at the critical point is not the $1/r$ form but $V_{screen}(r) \sim \frac{1}{r^{1+\eta_{\phi}}}$. 
Thus, the Coulomb interaction is screened but not completely in contrast to the Thomas-Fermi screening in metals.

\section*{Conclusion and Discussion}
We presented a controlled RG theory for the Coulomb interaction and disorder in Dirac semi-metals in three dimensions.
Fascinating interplay between the two driving forces induces a novel interacting quantum critical point 
between the disorder-dominant phase and the marginally-renormalized quantum critical phase.
The disorder generally enhances the density of states and tends to enhance the long-range Coulomb interaction. 
However, the long-range interactions would also be screened by ``more"
virtual particle-hole excitations along the way. The competition and balance between these effects eventually leads to 
a new interacting fixed point, where the NFL ground state arises. 
Using generalized forms of disorder distribution,
the NFL is accessed in a controlled way,
which has never been achieved in other disordered interacting systems.

Novel characteristics of the non-Fermi-liquid phase found here can be further emphasized by comparing  our theory
with others in literature. It is well understood that electrons in Dirac semi-metals  coupled to order parameters do not receive anomalous dimensions in four space-time \cite{zinnjustin}.
It is because spontaneous-symmetry-breaking transitions tend to open up { an} energy gap to 
maximize ``condensation'' energy, so at the critical point, 
electrons are about to feel the gap-opening related to suppressed density-of-states in the ordered phase. 
Even with strong anisotropy as in a recently discussed quantum phase transition between Weyl semi-metals and topological insulators, 
the long-range Coulomb interactions are screened too strongly and become irrelevant
\cite{yang}. 
Its ground state becomes another Slater-type product state.
Thus, we argue that the Dirac semi-metals with disorder and Coulomb interactions present
a unique setting for the NFL behaviors.

Our discussion on NFL can straightforwardly be applied to Weyl semi-metals suggested in pyrochlore iridates \cite{ashvin,kim} and 
topological quantum phase transitions \cite{shindou, goswami,fu} with a modified $N_f$.  
Future theoretical studies may include more comprehensive treatment of (small) doping and DC transport. 
Interplay between the NFL physics and rare region physics \cite{huse} would be 
another intriguing subject of study.

\bibliographystyle{naturemag}

{\small \subsection*{Acknowledgements}
We greatly appreciate discussions with L. Balents, L. Fu, V. Gurarie, A. Ludwig, L. Radzihovsky, J. Rau, D. T. Son, and  S. Syzranov.
We are particularly grateful to Igor Herbut for discussions about RG calculations.
EGM is grateful to the Boulder School 2014, where some part of this work was done.
YBK acknowledges the hospitality of the Aspen Center for Physics (NSF Grant No. PHYS- 1066293), where some part of this 
work was performed.
EGM is supported by the Kadanoff Center Fellowship and the MRSEC Program of the National Science Foundation under Award No. DMR 1121053.
 YBK is supported by NSERC, CIFAR, and Center for Quantum Materials at the University of Toronto.


%
%
%
%
%
%
%
%
%
%
%
%

%
%
%

\newpage

\noindent{\bf Supplementary Note 1}
\\
{\bf Feynman Diagrams and RG.}
In this section, we provide more information on the evaluations of the Feynman diagrams in Fig. 2 of the main-text.
The notation for the integrals used in the main text and below is 
\begin{eqnarray}
\int_{x , \tau} = \int d^3 x \int d \tau\quad, \quad \int_{k, \omega} = \int \frac{d^3 k}{(2\pi)^3} \int \frac{d \omega}{2 \pi}. \nonumber
\end{eqnarray}
The disorder Hamiltonian is written as 
\begin{eqnarray}
{H}_{imp}= \int_x r_{\kappa} \, \eta(x) \Psi^{\dagger}M^{\kappa}\Psi(x).
\end{eqnarray}
The matrices $M^{\kappa}$ correspond to different disorder types: $M^0=I$ for chemical potential disorder and $M^1$ with $\{\mathcal{H}_0, M^1 \}=0$ for mass gap disorder. 
Vector potential disorder is considered in the next section. 
We do not consider electrically charged disorders, so their coupling to the electric field is absent in the Hamiltonian. Higher order contribution is suppressed by the density fluctuation due to the charge conservation, so it does not contribute to beta functions.

As usual, the non-interacting Green's function is defined as
\begin{eqnarray}
&&G_f(k, i k_n)=\frac{1}{-i k_n + \mathcal{H}_0(k) } \equiv \sum_{\alpha=\pm} \frac{ {\rm P_{\alpha}(k)}}{- i k_n +\alpha v_F |k|} \nonumber \\
&& {\rm P_{\alpha}(k)} = |k, \alpha><k, \alpha|, \quad \mathcal{H}_0(k) |k, \alpha> = \alpha v_F |k| |k, \alpha.> \nonumber
\end{eqnarray} 
Note that ${\rm P_{\alpha}(k)}$ is a projection operator. 
This operator is not normalized, so the normalized Green's function has the form, 
\begin{eqnarray}
G^R_f(k, i k_n) = \sum_{\alpha=\pm} \frac{\mathcal{A} \,{\rm P_{\alpha}(k)}}{(- i k_n +\alpha v_F |k|^z)^{1-\eta_f}} \mathcal{G}(\frac{i k_n}{k^z}).
\end{eqnarray}
A dimensionless scaling function $\mathcal{G}(x)$ and non-universal constant $\mathcal{A}$ are introduced, and the anomalous dimensions ($z$ and $\eta_f$) are obtained below.

The electron self-energy from disorder (Coulomb) scattering is 
\begin{eqnarray}
\Sigma_{f}^{r,e}(k, i \omega) &=& r_{\kappa}^2\int_{q, i q_n} M^{\kappa} \frac{G_{f}(k+q, i q_n+i \omega_n)}{\Pi_{r,e}{(q, iq_n)}} M^{\kappa} \nonumber 
\end{eqnarray}
with $\Pi^{-1}_{e,r}$ being the disorder or Coulomb propagator. 
Note that $\Pi_e$ is frequency independent due to instantaneous Coulomb interaction and $\Pi_r$ has only zero-frequency component 
because it represents a quenched disorder.
The Coulomb potential self-energy is 
\begin{eqnarray}
\Pi_e &=& e^2 \int_{q,i q_{n}} {\rm Tr} \big[G_{f}(k+q, i k_n+i q_n)G_f(q, iq_n) \big]. \nonumber
\end{eqnarray}
We remark that the disorder propagator is non-local, so it is not renormalized by quantum corrections. 
Due to the non-renormalization, the replica index does not appear in our beta function calculation at the one-loop calculation, so our beta function is valid in the replica limit $R \rightarrow 0$. 

The disorder vertex correction (sum of the first and second diagrams in Fig. 2 (c))  is 
\begin{eqnarray}
V^{\kappa} &=& \int_{q, i q_n} M^{\kappa} G_f (q, i q_n) M^{\kappa} G_f (q, i q_n) M^{\kappa} \frac{r_a^2}{\Pi_r(q, iq_n)} \nonumber \\
&+& \int_{q, i q_n} G_f (q, i q_n) M^{\kappa} G_f (q, i q_n) \frac{-e^2}{\Pi_e(q, iq_n)}, \nonumber
\end{eqnarray}
and the Coulomb vertex correction (sum of  the third and fourth diagrams in Fig. 2 (c))  is 
\begin{eqnarray}
V_e &=& \int_{q, i q_n} M^{\kappa} G_f (q, i q_n) G_f (q, i q_n) M^{\kappa} \frac{r_a^2}{\Pi_r(q, iq_n)} \nonumber \\
&+& \int_{q, i q_n} G_f (q, i q_n) G_f (q, i q_n) \frac{-e^2}{\Pi_e(q, iq_n)}. \nonumber
\end{eqnarray}

Now it is straightforward to perform the standard RG analysis. 
The frequency dependent part of the electron self-energy is
\begin{eqnarray}
\Sigma_{f}^0(0, i \omega) 
&\equiv& \frac{{\rm Tr}( \Sigma_{f} \Gamma^0)}{{\rm Tr}(\Gamma^0)}= i \omega \int_q \frac{1}{ \omega^2 +E(q)^2} \frac{r_{\kappa}^2 }{q^{m_c - \epsilon}}. \nonumber
\end{eqnarray}
The divergent piece at the critical dimension $\epsilon=0$ with $m_c=1$ is 
\begin{eqnarray}
\Sigma_{f}^0(0, i \omega) \Big|_{div} = i \omega \frac{r_{\kappa}^2 }{2 \pi^2 v_F^2} \log(\Lambda) . \nonumber 
\end{eqnarray}
It is convenient to introduce the following dimensionless coupling constants, 
\begin{eqnarray}
g_1 = \frac{2 e^2}{ \pi v_F},\,\, g_{2,0} = \frac{r_{0}^2 \Lambda^{\epsilon} }{2 \pi^2 v_F^2 },\,\, g_{2,1} = \frac{r_{1}^2 \Lambda^{\epsilon} }{2 \pi^2 v_F^2 }. \nonumber
\end{eqnarray}
The disorder vertex correction from the disorder-fluctuation fields (Fig. 2 (c), the first diagram) is 
\begin{eqnarray}
V^{\kappa, r} &=& \int_{q, i q_n} M^{\kappa} G_f (q, i q_n) M^{\kappa} G_f (q, i q_n) M^{\kappa} \frac{r_a^2}{\Pi_r(q, iq_n)} \nonumber \\
&=&M^{\kappa} \int_q \frac{(-1)^{\kappa}}{E_q^2} \frac{1}{q} \nonumber \\
V^{\kappa, r}|_{div }&=& (-1)^{\kappa} M^{\kappa} \frac{r_{\kappa}^2}{2 \pi^2 v_F^2} \log (\Lambda). \nonumber 
\end{eqnarray}
Similarly, the Coulomb potential and the vertex corrections have logarithmically-divergent pieces. 

The summary of the divergences (for the purpose of the RG analysis) is  
\begin{eqnarray}
&& \Sigma_{f}^a(k,0)\big|_{div} = -\epsilon_a(k) \frac{g_1-(-)^{\kappa}g_{2 \kappa}}{3} \log(\Lambda) \nonumber \\
&& \Pi_e(q)\big|_{div} = \frac{q^2}{4 \pi}  (N_f c_{\varphi}) g_1 \log(\Lambda) \nonumber \\
&& V^{\kappa}\big|_{div} = \Big[(-)^{\kappa} g_{2 \kappa} +\frac{1-(-1)^{\kappa}}{2}\frac{g_1}{2} \Big] M^{\kappa}\log(\Lambda) \nonumber \\
&& V_e\big|_{div} = g_{2 \kappa} \log(\Lambda). \nonumber 
\end{eqnarray}
The dynamic critical exponent is $z=1-\frac{g_1}{3}+\frac{4 g_2}{3}$ and the anomalous dimension is $\eta_f = g_2$.
Then, the beta functions of the two coupling constants for the chemical potential disorder can be obtained from these 
informations and are given by
\begin{eqnarray}
&&\frac{d}{d l} g_1 = - (N_f c_{\varphi}+\frac{1}{3}) (g_1)^2+\frac{4}{3} g_1 g_{2,0} \nonumber \\
&&\frac{d}{d l} g_{2,0} = - \epsilon \,g_{2,0} + \frac{8}{3} (g_{2,0})^2 -\frac{2}{3} g_1 g_{2,0}, \nonumber
\end{eqnarray}
with a flavor number, $N_f$, and $c_{\varphi} = \frac{ {\rm Tr} (\Gamma^0) }{12}$. 

There are three fixed points, $(g_1, g_{2})= \big[O=(0,0),U= (0, \frac{3 \epsilon}{8})$, $ QC=(\frac{\epsilon}{2 N_f c_{\varphi}}, (3+\frac{1}{N_f c_{\varphi}}) \frac{\epsilon}{8}\big]$ for non-zero positive $\epsilon$. 
At $U$, the linearized beta functions are 
\begin{eqnarray}
&&\frac{d}{d l} \delta g_1 = \frac{ \epsilon}{2} \delta g_1\, , \quad \frac{d}{d l} \delta g_{2} = \epsilon \, \delta g_{2} -\frac{\epsilon}{4} \delta g_1, \nonumber
\end{eqnarray}
where $\delta g_{1,2} = g_{1,2}- g_{1,2}|_U$. It is manifest that the Coulomb coupling $g_1$ becomes relevant. The fixed point has two relevant operators, so it does not describe a usual phase transition. 

At $QC$, the most important feature is that the effective fine structure constant ($g_1$) is non-zero.
The Coulomb interactions and the associated correlations between electrons are significant in the ground state of the critical point and 
qualitatively different behaviors from the non-interacting ground state appear. 
To capture them, one can linearize the beta functions finding one positive eigenvalue, $\nu^{-1}\equiv (3+(N_f c_{\varphi})^{-1} + \sqrt{81+30/(N_f c_{\varphi}) + 1/(N_f c_{\varphi})^2}) \frac{\epsilon}{12}$ and one negative eigenvalue, $d_n= (3+(N_f c_{\varphi})^{-1} - \sqrt{81+30/(N_f c_{\varphi}) + 1/(N_f c_{\varphi})^2}) \frac{\epsilon}{12}$.  
As expected, the one relevant operator tunes the quantum phase transition. On the other hand, the other operator becomes irrelevant, which 
is a consequence of balancing the enhanced density of states and the Coulomb screening. 
For $\epsilon=1$ and $N_f=2$, we find $\nu=0.76$ , $z=1.5$, $\eta_f = 9/16$ and $\nu z =1.14$. 

Scaling dimensions of fermion bilinear operators can be evaluated by the standard operator insertion methods \cite{s_qft}. 
For example,  anomalous dimensions of the operator $\psi^{\dagger} \Gamma^a \psi$, $a=1,2,3$ can be evalutaed by 
two diagrams with the vertex $\Gamma^{a}$  similar to the (c) 1st and 2nd ones in the main-text. We find its scaling dimension is $[\psi^{\dagger} \Gamma^{a}\psi] = d+\frac{\epsilon}{2}$.

The crossover boundary is determined by comparing scaling dimensions of temperature and the relevant operator $g_2$ at $QC$, $T^* \sim |g_2 -g_{QC}|^{z\nu}$
,
\begin{eqnarray}
z \nu = \frac{12  (\frac{1}{\epsilon}+ \frac{1}{2})}{ (3+(N_f c_{\varphi})^{-1} + \sqrt{81+30/(N_f c_{\varphi}) + 1/(N_f c_{\varphi})^2}) }. \nonumber
\end{eqnarray}
The minimum numbers of $N_f$ and $c_{\varphi}$ for the Dirac semimals are $N_f=2$ and $c_{\varphi}=1/3$. 
Thus we have $z \nu \ge 1.14$ in the one-loop calculations.

Note that we have considered different disorder types separately. Due to different quantum corrections, the most relevant operator (chemical potential disorder) becomes dominant at low energy (See also \cite{s_sarma}).
To be explicit, the beta functions of the mass gap disorder are  
\begin{eqnarray}
&&\frac{d}{d l} g_1 = - (N_f c_{\varphi}+\frac{1}{3}) (g_1)^2+\frac{2}{3} g_1 g_{2,1} \nonumber \\
&&\frac{d}{d l} g_{2,1} = - \epsilon \,g_{2,1} - \frac{8}{3} (g_{2,1})^2 +\frac{1}{3} g_1 g_{2,1}. \nonumber
\end{eqnarray}
There is only one stable fixed point $(0,0)$ in this case.

\newpage
\noindent{\bf Supplementary Note 2}
\\
{\bf  Uncontrolled momentum shell RG with white noise.}

In this section, we perform the Wilsonian shell RG with the white noise disorder. Strictly speaking, this method is uncontrolled, 
but it can capture qualitatively similar RG flows as in the previous controlled calculation if careful evaluation is done. 

Since the electric charge is marginal, the main difference from the previous calculation comes from the disorder interaction. For example, the frequency dependence of the self-energy is 
\begin{eqnarray}
\Sigma_{f}^0(0, i \omega) 
= \int_{\Lambda e^{- dl}}^{\Lambda} \frac{i \omega \, r_{\kappa}^2}{ \omega^2 +E(q)^2} = i \omega \frac{r_{\kappa}^2 \Lambda}{2\pi^2 v_F^2} \,dl +O(dl^2).\nonumber
\end{eqnarray}
Following the same procedure, we find
\begin{eqnarray}
&& \Sigma_{f}^a(k,0) = -\epsilon_a(k) \frac{g_1}{3} dl +O(dl^2) \nonumber \\
&& \Pi_e(q) = \frac{q^2}{4 \pi}  (N_f c_{\varphi}) g_1 dl +O(dl^2) \nonumber \\
&& V^{\kappa} = \Big[(-)^{\kappa} g_{2 \kappa} +\frac{1-(-1)^{\kappa}}{2}\frac{g_1}{2} \Big] M^{\kappa}dl +O(dl^2) \nonumber \\
&& V_e = g_{2 \kappa}dl +O(dl^2). \nonumber 
\end{eqnarray}

The beta functions for the chemical potential disorder case is 
\begin{eqnarray}
&&\frac{d}{d l} g_1 = - (N_f c_{\varphi}+\frac{1}{3}) (g_1)^2+ g_1 g_{2,0} \nonumber \\
&&\frac{d}{d l} g_{2,0} = - \,g_{2,0} + 2 (g_{2,0})^2 -\frac{2}{3} g_1 g_{2,0}. \nonumber
\end{eqnarray}
Note that $\epsilon$ is replaced by $1$ in the first term of the second line, and the coefficients of $(g_{2,0})^2$ term is modified with the factor $4/3$, which comes from the fact that the electron-self energy does not receive momentum dependent corrections with the white-noise disorder. There are three fixed points : $O = (0,0)$, $D'=(0,\frac{1}{2})$, and $QC'=(\frac{1}{2 N_f c_{\varphi}}, \frac{1}{2}+\frac{1}{6 N_f c_{\varphi}})$.
Remark that the beta functions are consistent with previous calculations in literature for the two limits : $g_1 \rightarrow 0$  and $g_2 \rightarrow 0$. \cite{s_gurarie, s_hosur}
To compare with the Ref. \cite{s_hosur}, it is convenient to define $g_1= \frac{2 e^2}{\pi v_F}= \frac{2}{\pi}\alpha $ and use ${\rm Tr} (\Gamma_0) =2$.

We remark that the one-particle reducible diagrams in \ref{irrelevant} do not contribute to our beta function calculation. 
In addition to the reducibility, the self-energy diagram in \ref{irrelevant} (a) is independent of external frequency and momentum, 
so they do not renormalize the electron propagator. 
The related vertex corrections by operator insertions in \ref{irrelevant} (b) should vanish via the Ward identity
\begin{eqnarray}
q_{\mu} \Lambda^{\mu}(p+q,p) &=& -G^{-1}(p+q) +G^{-1} (p) \nonumber \\
q_{\mu} \Gamma^{\mu}(p+q,p) &=& \Sigma_f (p+q) - \Sigma_f(p), \quad G^{-1}(p) = p_{\mu} \gamma^{\mu} - \Sigma_f(p)
\end{eqnarray}
where $\Gamma^{\mu}(p,p')$ is the correction to the vertex function ($\Lambda^{\mu} =\gamma^{\mu}+ \Gamma^{\mu}$) with the bare vertex $\gamma^{\mu}$) in the relativistic notation $q_{\mu} = (q_n, \vec{q})$ \cite{s_qft}. Note that we focus on the density vertex $\Gamma^0(p,p')$ in the current work such as $V_e$ and $V^{\kappa}$ depending on the vertex types.
Thus, the reducible tadpole self-energies and related vertex corrections do not contribute to the beta functions. 
Notice, for example, that the exclusion of such reducible diagrams was discussed in 
a previous work\cite{s_herbut} in the context of bosonic theories and is the consequence of charge neutrality.
The non-contribution might be a source of the discrepancy between our work and the previous work\cite{s_goswami}.

The fixed points and RG flow structure in \ref{white} are basically the same as the one in the main text even though the small parameter in the beta function is absent in the white-noise case. 
For $N_f=2$ Dirac semimetals, we find $\nu=0.76$ , $z=1.5$, and $\eta_f =0.75$ for the shell RG. Note that the critical exponents are indeed similar to those of the $\epsilon$ method in the main text, which is another signature of validity of our results. 

To be self-contained, we consider a vector potential type disorder ($\psi^{\dagger} \Gamma^a \psi$). In the momentum shell RG procedure, only the disorder vertex correction is different from the previous calculations. We introduce one more dimensionless parameter, $g_{2,v} =\frac{r_{v}^2 \Lambda}{2 \pi^2 v_F^2 } $ to characterize the strength of the vector potential disorder. 
The vertex correction is 
\begin{eqnarray}
&& V^{a}_v = \Big[-\frac{g_{2,v}}{3} +\frac{g_1}{6} \Big] M^a dl +O(dl^2).
\end{eqnarray}
Note that its sign is the same as one of the mass gap disorder. It is because the vector potential term anticommutes with two Gamma matrices and commute with one Gamma matrix. 
Then, the beta funcitons are 
\begin{eqnarray}
&&\frac{d}{d l} g_1 = - (N_f c_{\varphi}+\frac{1}{3}) (g_1)^2+ g_1 g_{2,v} \nonumber \\
&&\frac{d}{d l} g_{2,v} = - \,g_{2,v} -\frac{8}{3} (g_{2,v})^2 -\frac{1}{3} g_1 g_{2,v}. \nonumber
\end{eqnarray}
As in the mass gap disorder, there is only one stable fixed point, $(0,0)$.

\newpage
\noindent{\bf Supplementary References}

\
\newpage
\begin{figure*}[t]
\centering
\includegraphics[width=16 cm]{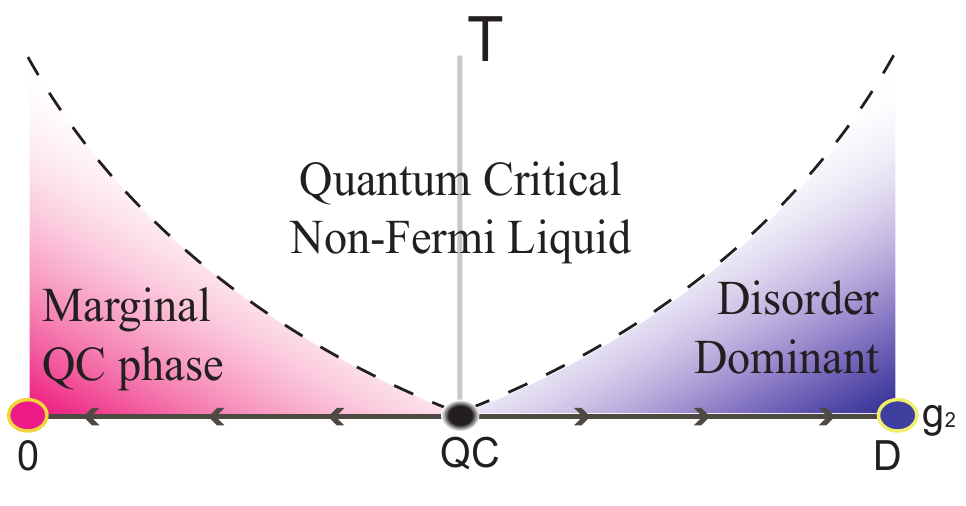}
\caption{
{\bf Quantum phase transition between the marginally-renormalized quantum critical phase $O$ (the Dirac semi-metal with
logarithmic corrections) and a disorder-dominant ground state ($D$). The horizontal (vertical) axis is for disorder strength (temperature). 
The ground state of the quantum critical point ($QC$) is a non-Fermi liquid and the finite-temprature quantum critical fan is unusually wide, which is determined by the dotted crossover line, $T^* \sim |g_2-g_{QC}|^{\nu z}$. For $\epsilon=1$ and $N_f=2$, we find $\nu=0.76$ , $z=1.5$, and $\nu z =1.14$. }
}
\label{phasediagram}
\end{figure*}

\newpage
\begin{figure*}[t]
\centering
\includegraphics[width=16 cm]{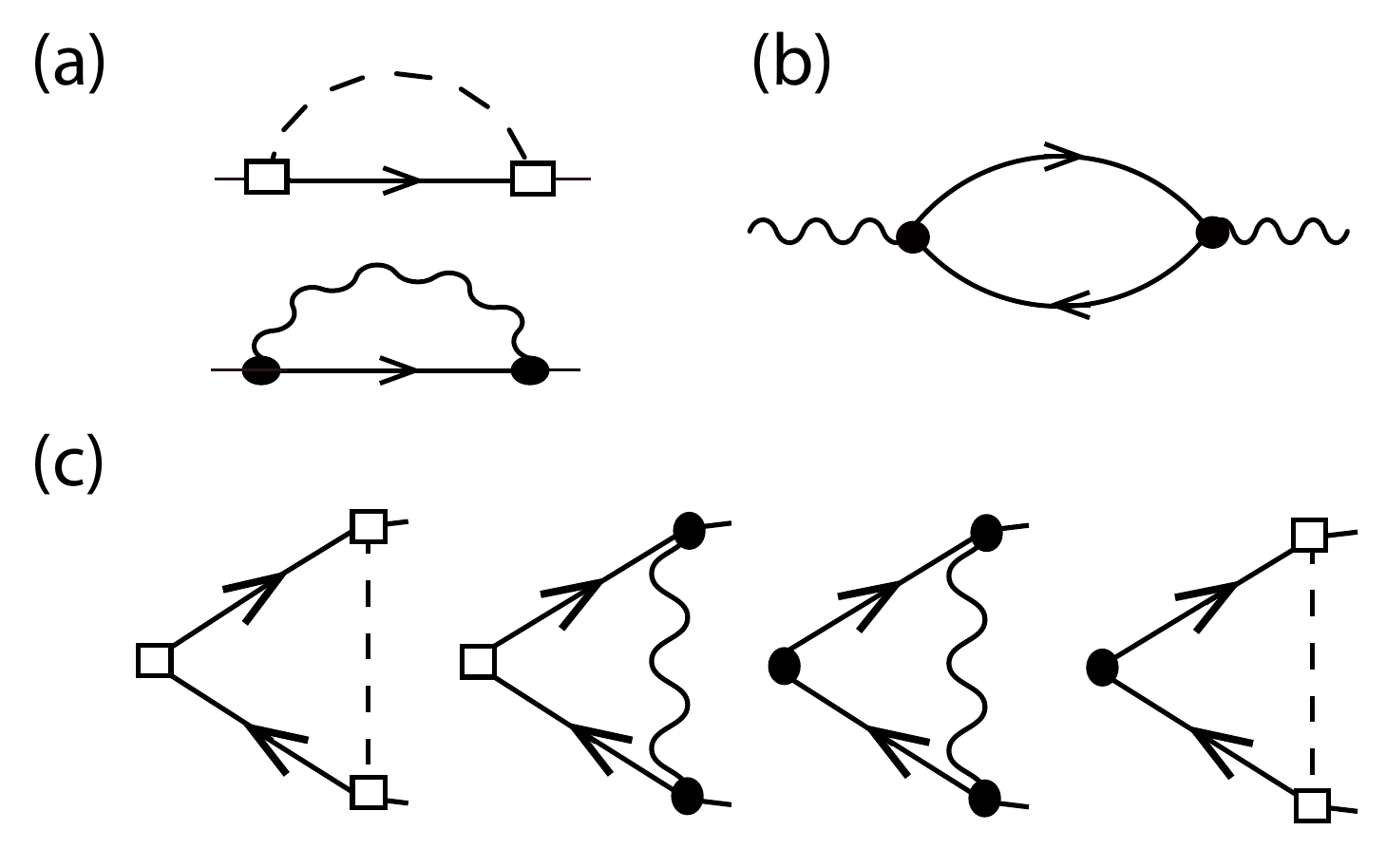}
\caption{
{\bf  Feynman diagrams used in the RG analysis. Solid, wavy, and dashed lines correspond to the electron, Coulomb interation, and disorder propagators. 
The filled circle (empty square) is for the Coulomb interaction (disorder) vertex.
(a) electron self-energy. (b) Coulomb potential self-energy. (c) vertex corrections. }
} \label{diagram}
\end{figure*}
\newpage
\begin{figure*}[t]
\centering
\includegraphics[width=16 cm]{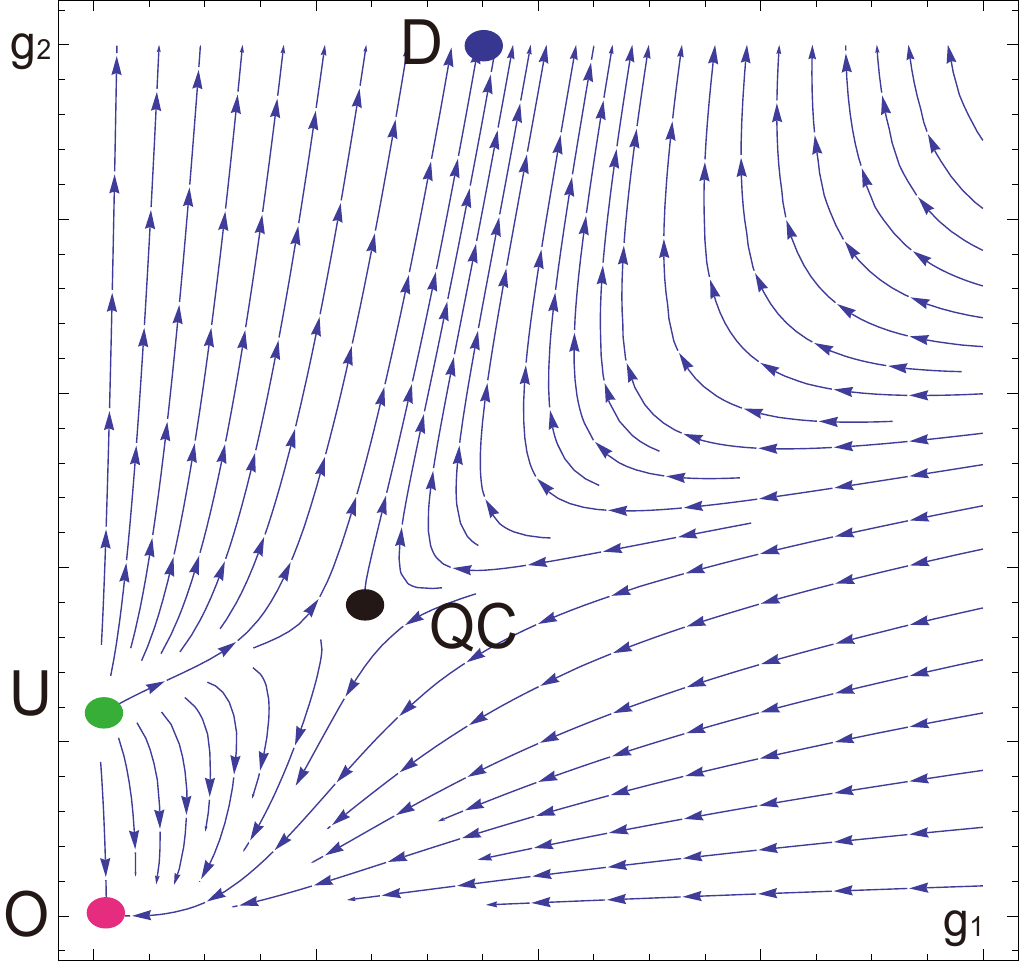}
\caption{
{\bf  RG flow diagrams. There are three fixed points, $\big( O= (0,0), U=(0, \frac{3 \epsilon}{8}), QC= (\frac{\epsilon}{2 N_f c_{\varphi}}, (3+\frac{1}{N_f c_{\varphi}}) \frac{\epsilon}{8}\big)$ in addition to the disorder-dominant ground state ($D$). 
The arrow flows toward the IR limit. The fixed point $QC$ describes the phase transition between $O$ and $D$, which is illustrated in Fig. 1. }
} \label{flowdiagram}
\end{figure*}

\
\newpage
\begin{figure*}[t]
\centering
\includegraphics[width=16 cm]{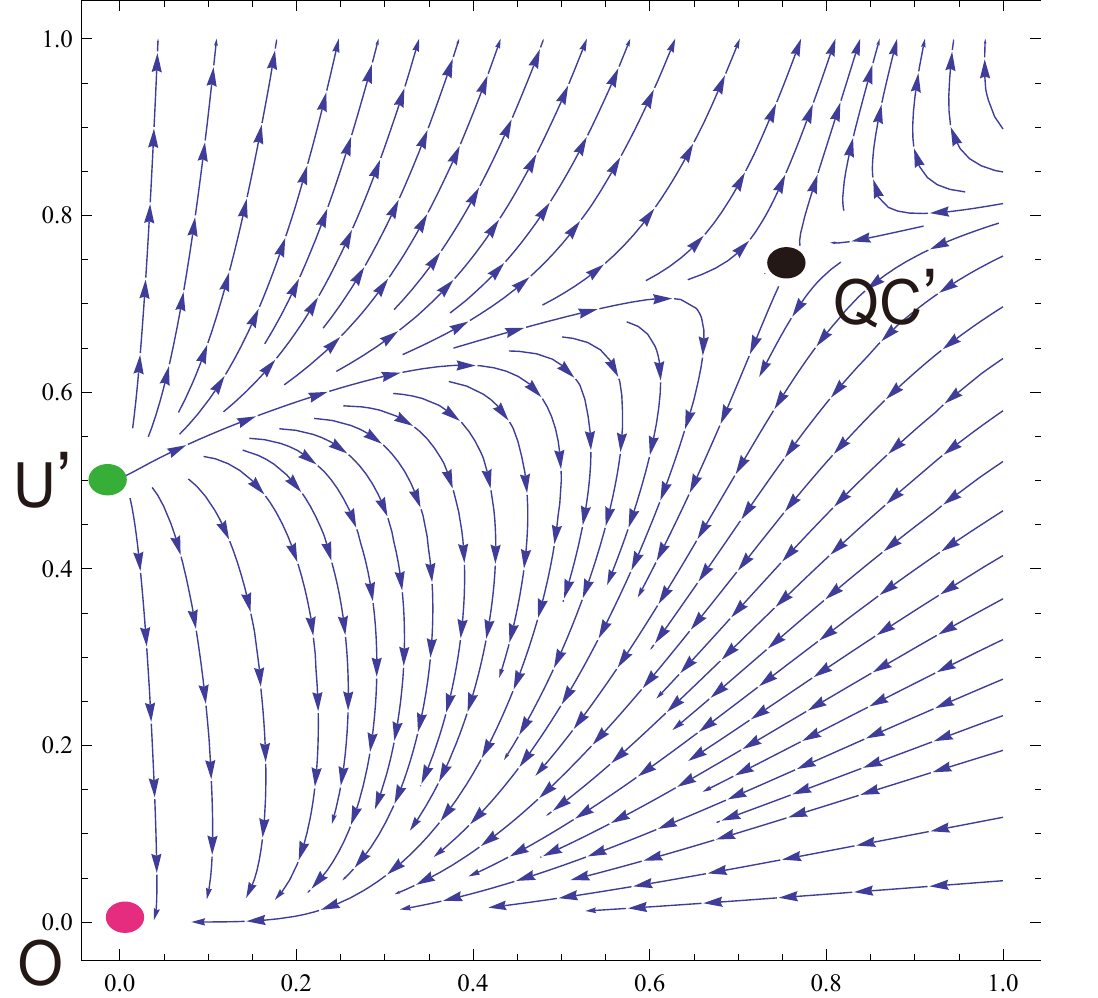}
\caption{
{\bf  RG flow of the uncontrolled white-noise calculation. The RG flow is basically the same as that of the $\epsilon$ expansion. There are three fixed points : $O = (0,0)$, $U'=(0,\frac{1}{2})$, and $QC'=(\frac{1}{2 N_f c_{\varphi}}, \frac{1}{2}+\frac{1}{6 N_f c_{\varphi}})$.  }
}
\label{white}
\end{figure*}

\
\newpage
\begin{figure*}[t]
\centering
\includegraphics[width=16 cm]{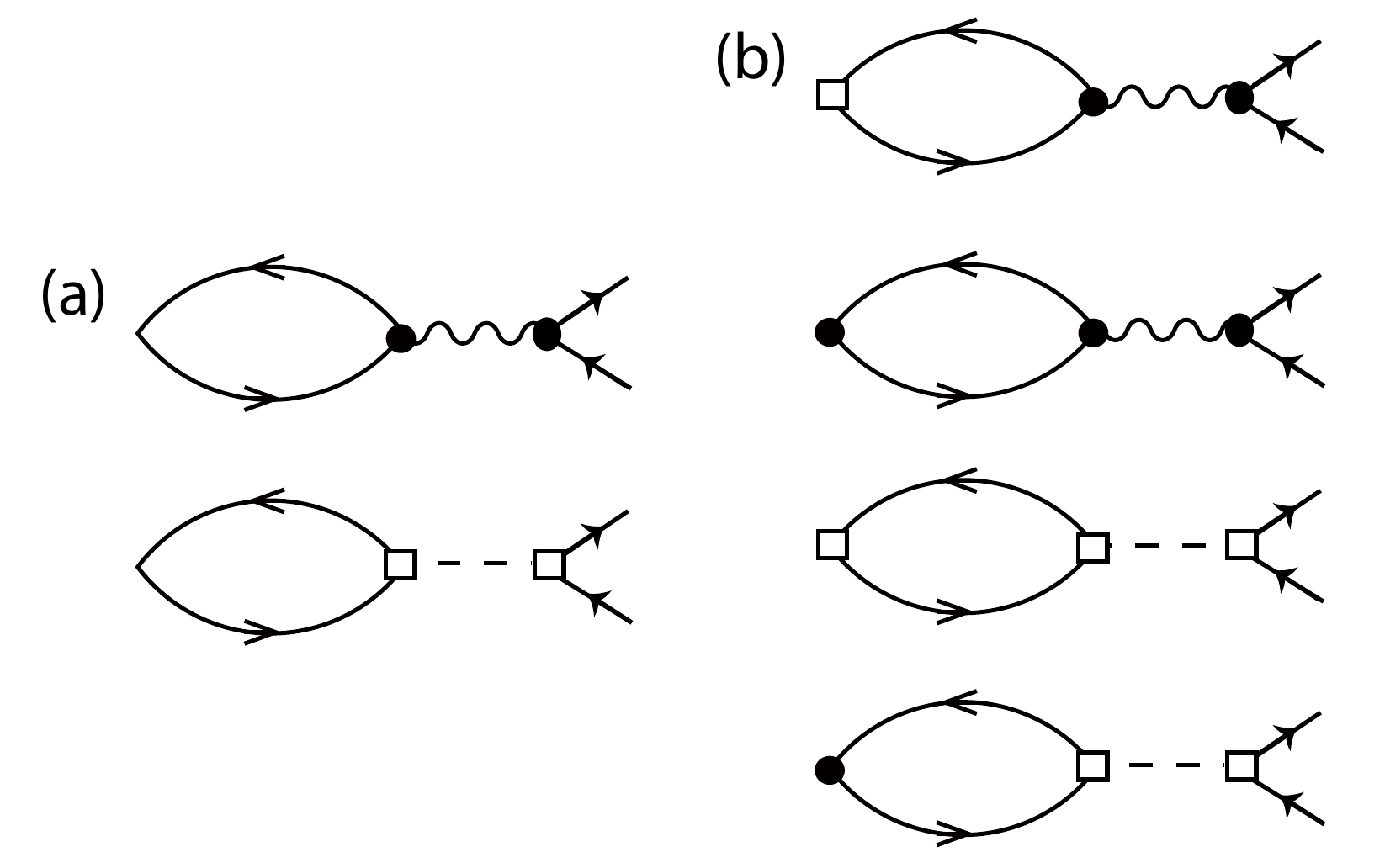}
\caption{
{\bf  Irrelevant diagrams for the beta functions. (a) the tadpole self-energy diagram (b) related vertex diagrams obtained by operator insertions. These are one-particle reducible  diagrams.  }
}
\label{irrelevant}
\end{figure*}

\end{document}